\documentclass[11pt,a4paper,english,preprint]{elsarticle}

\usepackage{amssymb}
\usepackage{amsmath}
\usepackage{graphicx}
\usepackage{babel}
\usepackage[T1]{fontenc}
\usepackage{hyperref}

\makeatletter
\newsavebox{\uuunit}
\sbox{\uuunit}
    {\setlength{\unitlength}{0.825em}
     \begin{picture}(0.6,0.7)
        \thinlines
        \put(0,0){\line(1,0){0.5}}
        \put(0.15,0){\line(0,1){0.7}}
        \put(0.35,0){\line(0,1){0.8}}
       \multiput(0.3,0.8)(-0.04,-0.02){12}{\rule{0.5pt}{0.5pt}}
     \end {picture}}
\newcommand {\unity}{\mathord{\!\usebox{\uuunit}}}
\makeatother

\begin{document}

\title{Equivalence between supersymmetric self-dual and Maxwell-Chern-Simons models coupled to a matter spinor superfield}

\author[abc]{A. F. Ferrari}
\ead{alysson.ferrari@ufabc.edu.br}

\author[usp]{M. Gomes}
\ead{mgomes@fma.if.usp.br}

\author[rn,usp]{A. C. Lehum}
\ead{lehum@fma.if.usp.br}

\author[pb]{J. R. Nascimento} 
\ead{jroberto@fisica.ufpb.br}

\author[pb]{A. Yu. Petrov}
\ead{petrov@fisica.ufpb.br}

\author[usp]{A. J. da Silva}
\ead{ajsilva@fma.if.usp.br}

\address[abc]{Centro de Ci\^encias Naturais e Humanas, Universidade Federal do ABC, Rua Santa Ad\'elia, 166, 09210-170, Santo Andr\'e, SP, Brazil}

\address[usp]{Instituto de F\'\i sica, Universidade de S\~ao Paulo\\
Caixa Postal 66318, 05315-970, S\~ao Paulo, SP, Brazil}

\address[rn]{Escola de Ci\^encias e Tecnologia, Universidade Federal do Rio Grande do Norte\\
Caixa Postal 1524, 59072-970, Natal, RN, Brazil}

\address[pb]{Departamento de F\'{\i}sica, Universidade Federal da Para\'{\i}ba\\
 Caixa Postal 5008, 58051-970, Jo\~ao Pessoa, Para\'{\i}ba, Brazil}

\begin{abstract}
We study the duality of the supersymmetric self-dual and Maxwell-Chern-Simons theories coupled to a fermionic matter superfield, using a master action. This approach evades the difficulties inherent to the quartic couplings that appear when matter is represented by a scalar superfield. The price is that the spinorial matter superfield represents a unusual supersymmetric multiplet, whose main physical properties we also discuss.
\end{abstract}


\maketitle

\section{Introduction}

Duality is an important phenomenon in quantum field theory allowing to relate two different theories. One example in  $(2+1)D$~\cite{Townsend:1983xs,nieu} is the equivalence between the self-dual (SD) model, which does not possess gauge invariance, and the gauge-invariant Maxwell-Chern-Simons (MCS) model~\cite{Deser:1981wh}. 
Different aspects of this equivalence were studied in the literature, see for example~\cite{Karlhede:1986qd,Fradkin:1994tt,Bralic:1995ip,Banerjee:1995yf,Banerjee:1996sp,Gomes:1997mf,Anacleto:2001rp,Minces:1999tp}. The most important results of these papers were to establish the mapping between a massive Thirring model and the Maxwell-Chern-Simons theory, and between the self-dual model and the Maxwell-Chern-Simons theory. 
The equivalence was also studied in the supersymmetric counterparts of the SD and MCS models, both in the free case~\cite{Karlhede:1986qf} as well as in the presence of interactions with a scalar matter superfield~\cite{Ferrari:2006vy}. However, as we will argue shortly, there remains some delicate intricacies which motivated us to reexamine the duality in the supersymmetric case.

In the present decade, a considerable interest has been devoted to the study of field theories in noncommutative spacetime and the possibility of Lorentz symmetry violation, mainly due to their relevance to  quantum gravity. In this context, the duality in a noncommutative spacetime was considered in~\cite{Gomes:2008pi}, and in the presence of Lorentz violation in~\cite{Furtado:2008gs}.

The duality between the models can in principle be proved within two frameworks. The first of them is the gauge embedding method~\cite{Anacleto:2001rp,Ferrari:2006vy}, whose essence consists in the extension of the self-dual model to a gauge theory by adding to its Lagrangian carefully chosen terms that vanish on-shell. The equivalence of the resulting gauge model and the starting SD theory can be seen by comparing their equations of motion, and can also be tested at the quantum level. The second framework is the master action method, used for example in~\cite{nieu,Gomes:1997mf}, based on some primordial action (the master action) involving both the MCS and SD fields, coupled to some matter. Integration of this master action over the MCS field yields the SD action, whereas integration over the SD produces the MCS action, with appropriate couplings to the matter in both cases. Proceeding one step further, one can integrate over the remaining SD field in the first case, or over the MCS field in the second, finding the same effective self-interaction for the matter in both situations.

When the SD field is coupled to a bosonic matter, one complication arises, in the sense that the model is actually equivalent to a ``modified'' MCS theory, with a field-dependent factor in front of the Maxwell term~\cite{Anacleto:2001rp}. The source of this complication is essentially the appearance of quartic vertices involving the matter and the vector fields. When considering the duality in the supersymmetric case, the most natural matter supermultiplet is represented by a scalar superfield, which also couples to the vector (fermionic) superfields with a quartic vertex, so the same difficulty arises: the supersymmetric SD model is equivalent to a modified MCS theory~\cite{Ferrari:2006vy}. The presence of the quartic vertices also precludes an extension of the proof of the duality for noncommutative theories (which, however, have been studied in the context of the Seiberg-Witten map, see for example~\cite{nccs,Harikumar:2005ry}). One might wonder whether an interaction with a fermionic superfield, which does not induce a quartic vertex in the classical action, could make the study of the duality more transparent, and the aim of this work is to show that this is so, at least in the commutative case. The price to pay is that the fermionic matter superfield we have to introduce in such a study describes a non-minimal supersymmetric multiplet, involving four bosonic and four fermionic degrees of freedom.

The structure of this work looks as follows. In Section~\ref{classical}, we present the master action, and use the equations of motion to establish the duality at the classical level. In Section~\ref{quantum}, we study the duality at the quantum level, by inspecting the generating of the SMCS and SSD theories. All this work is made for quite general couplings; some particular cases are discussed in Section~\ref{instances}. In Section~\ref{matter}, the physical content of the fermionic matter superfield introduced by us is made explicit. In the Summary, the results are discussed; in particular, we comment on the possible extension of our work to the noncommutative spacetime.

\section{The duality at the classical level}\label{classical}

As a first step, we introduce the following master Lagrangian describing the interaction of a spinorial matter superfield $\Psi^\alpha$ with the spinor superfields $f_{\alpha}$ (which will be further identified with the self-dual superfield) and $A_{\alpha}$ (which will be further identified with the Maxwell-Chern-Simons superfield),
\begin{eqnarray}\label{eq1}
\mathcal{L}_{\rm master}=-\frac{m^2}{2}f^{\alpha}f_{\alpha}+m~f^{\alpha}W_{\alpha}
+\frac{m}{2}A^{\alpha}W_{\alpha}
+k^{\alpha}f_{\alpha}+j^{\alpha}A_{\alpha}+\mathcal{L}_{M}(\Psi)~,
\end{eqnarray}

\noindent
where $\mathcal{L}_M(\Psi)$ is the quadratic Lagrangian for the spinor matter superfield  $\Psi^{\alpha}$; $j^{\alpha}$ and $k^{\alpha}$ are currents depending on this superfield. Explicit forms for $\mathcal{L}_M(\Psi)$ and the currents will be presented later, at the moment, we can say that $j^{\alpha}$ is necessarily conserved ($D_{\alpha}j^{\alpha}=0$) due to gauge invariance. Here $W_{\alpha}\equiv\frac{1}{2}D^{\beta}D_{\alpha}A_{\beta}$ is the gauge invariant superfield strength constructed from the superfield $A_{\alpha}$. The Lagrangian $\mathcal{L}_{\rm master}$ is the natural superfield generalization of the one used in~\cite{Gomes:1997mf}, with the notations and conventions of~\cite{Gates:1983nr}. 

The equations of motion for the $A^\alpha$ and $f^\alpha$ superfields derived from Eq.~(\ref{eq1}) can be used to obtain the duality at the classical level. Varying the action $\int d^5 z\,\mathcal{L}_{\rm master}$ with respect to $f^\alpha$ we obtain,
\begin{equation}\label{ident1}
f_{\alpha}  =\frac{1}{m^{2}}k_{\alpha}+\frac{1}{m}W_{\alpha}\,
\end{equation}

\noindent
which, inserted in Eq.~(\ref{eq1}), yields $\mathcal{L}_{\rm master} = \mathcal{L}_{\rm SMCS}$, with
\begin{eqnarray}\label{eq7a}
\mathcal{L}_{\rm SMCS}&=&\frac{1}{2}W^{\alpha}W_{\alpha}
+\frac{m}{2}A^{\alpha}W_{\alpha}-\frac{\alpha}{4}(D^{\alpha}A_{\alpha})^2 \nonumber\\
&&+\left(j^{\alpha}+\frac{1}{2m}D^{\beta}D^{\alpha}k_{\beta}\right)A_{\alpha}
+\frac{1}{2m^2}k^{\alpha}k_{\alpha}+\mathcal{L}_{M}(\Psi)~.
\end{eqnarray}

\noindent
This last Lagrangian describes the supersymmetric Maxwell-Chern-Simons (SMCS) field coupled to the matter through the ``minimal'' coupling $A^{\alpha}j_{\alpha}$, plus a ``magnetic'' coupling $\frac{1}{2m}A^{\alpha}D^{\beta}D_{\alpha}k_{\beta}=\frac{1}{m}W^\alpha k_\alpha$, and a Thirring-like self-interaction $\frac{1}{2m^2}k^{\alpha}k_{\alpha}$ of the spinorial matter superfield.

Varying the master action with respect to $A^\alpha$ provides us with
\begin{equation}
\label{ident2}
W_\alpha  + \Omega_\alpha + j_{\alpha} = 0\,,
\end{equation}

\noindent
where $\Omega^{\alpha}\equiv (1/2)D^{\beta}D^{\alpha}f_{\beta}$. At this point, we recall the projectors on the transversal and longitudinal parts of a fermionic superfield $\eta^\alpha$,
\begin{equation}
\eta_{\parallel}^{\alpha}=-D^{\alpha}D^{\beta}\frac{1}{2D^{2}}\eta_{\beta}\quad;\quad \eta_{\perp}^{\alpha}=D^{\beta}D^{\alpha}\frac{1}{2D^{2}}\eta_{\beta}\,,\label{eq:project}
\end{equation}

\noindent
so that $D^\alpha \, \eta_\alpha^{\perp} = 0$. The explicit form of the transversal projector in Eq.~(\ref{eq:project}) allows us to rewrite Eq.~(\ref{ident2}) as
\begin{equation}
\label{ident3}
A_{\alpha}^{\perp}  =-f_{\alpha}^{\perp}-\frac{1}{mD^{2}}j_{\alpha}\,.
\end{equation}

\noindent
Substituting Eqs.~(\ref{ident3}) and ~(\ref{ident2}) into the master Lagrangian, and taking into account that, if $\eta^\alpha$ is transversal, $\eta^\alpha \xi_\alpha = \eta^\alpha \xi^\perp_\alpha$ for any $\xi_\alpha$, we obtain $\mathcal{L}_{\rm master} = \mathcal{L}_{\rm SSD}$, with
\begin{eqnarray}\label{eq9a}
\mathcal{L}_{\rm SSD}=-\frac{m}{2}f^{\alpha}\Omega_{\alpha}
-\frac{m^2}{2}f^{\alpha}f_{\alpha}
+\left(k^{\alpha}-j^{\alpha}\right)f_{\alpha}
-\frac{1}{2}j^{\alpha}\frac{1}{mD^2}j_{\alpha}
+\mathcal{L}_{M}(\Psi)~.
\end{eqnarray}

\noindent
This Lagrangian describes the dynamics of a supersymmetric Self-Dual (SSD) superfield which, besides of the ``minimal'' coupling to the current $k_{\alpha}$, is also coupled in a nonlocal way to the current $j_{\alpha}$. Moreover, a nonlocal Thirring-like term for the $j_{\alpha}$ shows up.

Classically, the Lagrangians in Eqs.~(\ref{eq7a}) and (\ref{eq9a}) are equivalent, thus establishing the duality between these SMCS and SSD models at the level of equations of motion. Indeed, we can find an explicit mapping between the superfields and currents of the SMCS theory to their counterparts in the SSD model, such that the corresponding equations of motion are mapped one to the other. The equations of motion derived from the SSD Lagrangian in Eq.~(\ref{eq9a}) can be cast as
\begin{equation}
m\Omega_{\alpha}+m^{2}f_{\alpha}+j^{\alpha}-k^{\alpha}=0\,,\label{eqq:2}
\end{equation}

\noindent
and
\begin{equation}
\frac{\delta}{\delta\Psi^\beta}\int d^5 z \mathcal{L}_{M} +
\frac{\partial j^{\alpha}} {\partial\Psi^\beta} \left(-f_{\alpha}^{\perp}-\frac{1}{mD^{2}}j_{\alpha}\right)+
\frac{\partial k^{\alpha}}{\partial \Psi^\beta} f_{\alpha}=0\,.\label{eqq:3}
\end{equation}

\noindent
Using the projection operators in Eq.~(\ref{eq:project}), we split Eq.~(\ref{eqq:2}) in the longitudinal,
\begin{equation}
m^{2}f_{\alpha}^{\parallel}=k_{\alpha}^{\parallel}\,,\label{eqq:2a}
\end{equation}

\noindent
and transversal parts,
\begin{equation}
m\Omega_{\alpha}^{\perp}+m^{2}f_{\alpha}^{\perp}+j_{\alpha}-k_{\alpha}^{\perp}=0\,.\label{eqq:2b}
\end{equation}

\noindent
Hereafter, we omit the $\perp$ in the current $j$ since we know it is always transversal. We see that the longitudinal part of $f$ is not dynamical, but algebraically related to the longitudinal part of $k_{\alpha}$. 

The equations of motion derived from the SMCS Lagrangian in Eq.~(\ref{eq7a}) read,
\begin{equation}
\frac{1}{2}D^{\beta}D_{\alpha}W_{\beta}+mW_{\alpha}+j_{\alpha}+\frac{D^{2}}{m}k_{\alpha}^{\perp}=0\,,\label{eqq:5}
\end{equation}

\noindent
and
\begin{equation}
\frac{\delta}{\delta\Psi^\beta}\int d^5 z \mathcal{L}_{M} +
\frac{\partial j^{\alpha}}{\partial\Psi^\beta} A_{\alpha}^{\perp}+
\frac{\partial k^{\alpha}}{\partial\Psi^\beta}
\left(\frac{1}{m^{2}}k_{\alpha}+\frac{1}{m}W_{\alpha}\right)=0\,.
\label{eqq:13}
\end{equation}

\noindent
All terms in Eq.~(\ref{eqq:5}) are transversal. Since $A^\alpha$ is a gauge superpotential, under a gauge transformation $\delta A^\alpha = D^\alpha K$, the transversal part $A_{\perp}^{\alpha}$ is invariant, while $\delta A_{\parallel}^{\alpha}=D^{\alpha}K$. Hence, the equations of motion involves the transversal (gauge invariant) part of $A^\alpha$; its longitudinal part is only constrained by the gauge fixing condition we will have to impose to quantize the theory~\cite{foot1}. 

Comparing Eqs.~(\ref{eqq:3}) and (\ref{eqq:13}), we conclude that one equation is mapped to the other by means of the equations of motion in Eqs.~(\ref{ident1}) and~(\ref{ident3}), so those furnish the identification we were looking for. Taking the longitudinal part of Eq.~(\ref{ident1}), we re-obtain Eq.~(\ref{eqq:2a}). Also, considering the transversal part of Eq.~(\ref{ident1}),
\begin{equation}
-f_{\alpha}^{\perp}+\frac{1}{m^{2}}k_{\alpha}^{\perp}+\frac{1}{m}W_{\alpha}^{\perp}
=0\,,\label{eqq:18}
\end{equation}

\noindent
and replacing $A_{\alpha}^{\perp}$ using Eq.~(\ref{ident3}), we re-obtain Eq.~(\ref{eqq:2b}). Finally, we can map the equation of motion for the SSD field in Eq.~(\ref{eqq:2b}) into the equation of motion for the transversal component of MCS superfield. Indeed, starting from
Eq.~(\ref{eqq:2b}), and substituting $f^{\perp}$ using Eq.~(\ref{ident3}),
we have
\begin{align}
m\Omega_{\alpha}^{\perp} & +m^{2}f_{\alpha}^{\perp}+j_{\alpha}-k_{\alpha}^{\perp}=\nonumber \\
= & -mW_{\alpha}^{\perp}-m^{2}A_{\alpha}^{\perp}-
\frac{m}{D^{2}}j_{\alpha}-k_{\alpha}^{\perp} = 0\,.\label{eqq:20}
\end{align}

\noindent
Applying $\frac{1}{2}D^{\alpha}D_{\beta}$ to this equation, we obtain
\begin{equation}
m\left[\frac{1}{2}D^{\alpha}D_{\beta}W_{\alpha}^{\perp}
+mW_{\beta}^{\perp}+\, j_{\alpha}+\frac{D^{2}}{m}k_{\alpha}^{\perp}\right]\,=\,0\,,\label{eqq:21}
\end{equation}

\noindent
which is equivalent to Eq.~(\ref{eqq:5}). 

In summary, the transversal (gauge invariant) part of the SMCS superfield can be mapped to the transversal part of SSD superfield. No relation exists between their longitudinal parts, however. In the SSD model, $f^\parallel$ is algebraically related to $k^\parallel$, while in the SMCS model, the longitudinal (gauge dependent) part of $A^\alpha$ is not coupled to other fields or currents, being constrained only by the choice of the gauge fixing.

\section{The duality at the quantum level}\label{quantum}

Having discussed the duality between $\mathcal{L}_{\rm SMCS}$ and $\mathcal{L}_{\rm SSD}$ at the level of equations of motion, we can now investigate whether this duality exists at the quantum level, by comparing the corresponding generating functionals. We will see that both theories lead to the same generating functional for the $\Psi^\alpha$ superfield, and in this sense we will say that $\mathcal{L}_{\rm SMCS}$ and $\mathcal{L}_{\rm SSD}$, as given in Eqs.~(\ref{eq7a}) and~(\ref{eq9a}), are quantum equivalent.

To this end, we have to include in the master Lagrangian a gauge fixing term, so that we can find a propagator for the SMCS superfield. We consider, then, the master generating functional,
\begin{align}\label{eq2}
Z(k^{\alpha},j^{\alpha},\Psi^{\alpha})& = \mathcal{N}
\int  \mathcal{D}f_{\alpha}\,\mathcal{D}A_{\alpha} \times \nonumber\\
&\times \exp \, i \int {d^5z} \left\{  \mathcal{L}_{\rm  master}(f^{\alpha},A^{\alpha},\Psi^{\alpha},j^{\alpha},k^{\alpha})
-\frac{\alpha}{4}(D^{\alpha}A_{\alpha})^2 \right\}~,
\end{align}

\noindent
where $\mathcal{N}$ is a field independent normalization factor. We will further need the formula for the Gaussian path integral over a Grassmannian field $X_{\alpha}$, 
\begin{eqnarray}\label{eq3}
\int~\mathcal{D}X_{\alpha}~
\exp\left\{ \,i \left[ \frac{1}{2}X^{\alpha}{\mathcal{O}_{\alpha}}^{\beta}X_{\beta}
+J^{\alpha}X_{\alpha}\right] \right\} \,=\, 
\exp\left\{ -i\left[\frac{1}{2}J^{\alpha}{(\mathcal{O}^{-1})_{\alpha}}^{\beta}J_{\beta}
\right]\right\},
\end{eqnarray}

\noindent 
up to a factor depending on ${\rm det}~{\mathcal{O}_{\alpha}}^{\beta}$, which will be irrelevant in this work, and omitting the proper superspace integrations in the exponents.

By means of Eq.~(\ref{eq3}), we can perform the functional integration in Eq.~(\ref{eq2}) over the superfield $f^{\alpha}$, with 
\begin{eqnarray}\label{eq4}
{({\mathcal{O}^{-1}_1})_{\alpha}}^{\beta}=-\frac{1}{m^2}{\delta_{\alpha}}^{\beta}~,
\end{eqnarray}

\noindent
and we end up with,
\begin{eqnarray}\label{eq5}
Z(k^{\alpha},j^{\alpha},\Psi^{\alpha})&=& \mathcal{N} \int \mathcal{D}A_{\alpha}\,\exp\Big{\{}i\int\!{d^5z}\Big[\frac{1}{2}(m~W^{\alpha}+k^{\alpha})\frac{{\delta_{\alpha}}^{\beta}}{m^2}(m~W_{\beta}+k_{\beta})\nonumber\\
&+&\frac{m}{2}A^{\alpha}W_{\alpha}+j^{\alpha}A_{\alpha}-\frac{\alpha}{4}(D^{\alpha}A_{\alpha})^2+\mathcal{L}_{M}\Big]\Big{\}}~,
\end{eqnarray}

\noindent 
which, after an integration by parts, can be cast as
\begin{equation}\label{eq6}
Z(k^{\alpha},j^{\alpha},\Psi^{\alpha})=\mathcal{N}
\int \mathcal{D}A_{\alpha}\,
\exp \, i \int {d^5z} \left\{ \mathcal{L}_{\rm SMCS} 
-\frac{\alpha}{4}(D^{\alpha}A_{\alpha})^2 \right\}~,
\end{equation}

\noindent 
where $\mathcal{L}_{\rm SMCS}$ is the SMCS Lagrangian we found in Eq.~(\ref{eq7a}).

To integrate the generating functional in Eq.~(\ref{eq2}) over $A_{\alpha}$, we use the inverse of the quadratic part in $A^{\alpha}$ of the master Lagrangian in Eq.~(\ref{eq1}), including the gauge-fixing term,
\begin{eqnarray}\label{eq8}
{(\mathcal{O}^{-1}_2)_{\beta}}^{\gamma}=\frac{1}{2}\Big(\frac{D^{\gamma}D_{\beta}}{m\Box}
+\frac{D_{\beta}D^{\gamma}}{\alpha\Box}\Big)~.
\end{eqnarray}

\noindent
Using Eq.~(\ref{eq3}), the functional integration in Eq.~(\ref{eq2}) over $A^{\alpha}$ can be performed, arriving at the following generating functional for the $f^{\alpha}$ and matter superfields,
\begin{eqnarray}\label{eq9}
Z(k^{\alpha},j^{\alpha},\Psi^{\alpha})&=&\int \mathcal{D}f_{\alpha} \,\exp\left\{
i\int {d^5z}~\mathcal{L}_{\rm SSD}\right\}~,
\end{eqnarray}

\noindent 
where $\mathcal{L}_{\rm SSD}$ is the SSD model defined in Eq.~(\ref{eq9a}).

To complete the proof of the equivalence of the SMCS and the SSD theories, we integrate the generating functionals in Eq.~(\ref{eq6}) over $A_{\alpha}$ and Eq.~(\ref{eq9}) over $f^{\alpha}$. The relevant propagators are
\begin{subequations}\label{props}
\begin{eqnarray}
{(\mathcal{O}_{SMCS}^{-1})_{\beta}}^{\alpha}&=&\frac{1}{2}\Big[\frac{D^{\alpha}D_{\beta}}{\Box(D^2+m)}
+\frac{1}{\alpha}\frac{D_{\beta}D^{\alpha}}{\Box}\Big]~,\label{prop1}\\
{(\mathcal{O}_{SSD}^{-1})_{\beta}}^{\alpha}&=&\frac{1}{2}\Big[\frac{D_{\beta}D^{\alpha}}{m^2D^2}
-\frac{1}{m}\frac{D^{\alpha}D_{\beta}}{D^2(D^2+m)}\Big]~ \nonumber \\
&=& \frac{1}{2m^2} \left[ \frac{D^\alpha D_\beta}{D^2+m} - 2 \delta^\alpha_{\,\,\beta}  \right]
\label{prop2} \,.
\end{eqnarray}
\end{subequations}

\noindent
The integration over $A^{\alpha}$ and $f^{\alpha}$ respectively in Eq.~(\ref{eq6}) and Eq.~(\ref{eq9}) results in the same effective Lagrangian,
\begin{eqnarray}\label{eq10}
\mathcal{L}_{eff}&=&-\frac{1}{4m^2}j^{\alpha}\Big(\frac{D^{\beta}D_{\alpha}}{D^2+m}-2{\delta_\alpha}^{\beta}\Big)j_{\beta}
-\frac{1}{2m}j^{\alpha}\frac{1}{D^2}j_{\alpha}
+\frac{1}{m}k^{\alpha}\frac{1}{D^2+m}j_{\alpha}\nonumber\\
&-&\frac{1}{4m^2}k^{\alpha}\Big(\frac{D^{\beta}D_{\alpha}}{D^2+m}-2{\delta_\alpha}^{\beta}\Big)k_{\beta}
+\mathcal{L}_{M}(\Psi)~,
\end{eqnarray}

\noindent
as it should. This ensures the quantum equivalence between the two models, irrespective of the choice of the currents $j^{\alpha}$ and $k^{\alpha}$ (whereas $D^\alpha\,j_\alpha=0$). 

One last note before closing this section. The physical content of the model in Eq.~(\ref{eq9}) can be more clearly seen by means of the field redefinition,
\begin{equation}
A_\alpha \, = \, B_\alpha - f_\alpha \,
\end{equation}

\noindent
which allows us to rewrite Eq.~(\ref{eq1}), apart from surface terms, as
\begin{align}\label{eq1eq}
\mathcal{L}_{\rm master}=&-\frac{m^2}{2}f^{\alpha}f_{\alpha}
-\frac{m}{2}~f^{\alpha}\Omega_{\alpha}+\frac{m}{2} B^{\alpha} \left( \frac{1}{2}D^{\beta}D_{\alpha}B_{\beta} \right) \nonumber\\
&+\left(k^{\alpha} - j^\alpha\right)f_{\alpha}+j^{\alpha}B_{\alpha}
+\mathcal{L}_{M}(\Psi)~,
\end{align}

\noindent
From Eq.~(\ref{eq1eq}), we see that $\mathcal{L}_{\rm master}$ describes a propagating field governed by a Self-Dual Lagrangian, together with a pure topological Chern-Simons field $B^{\alpha}$.

We can use the propagators in Eqs.~(\ref{props}) to find the superfields $f^{\alpha}$ and $B^{\alpha}$ in terms of the corresponding sources,
\begin{subequations}\label{src}
\begin{eqnarray}
f_\beta &=& - {(\mathcal{O}_{SSD}^{-1})_{\beta}}^{\alpha} \left(k_\alpha - j_\alpha \right) \,, \label{src1} \\
B_\beta &=& - {(\mathcal{O}_{SMCS}^{-1})_{\beta}}^{\alpha} j_\alpha \,. \label{src2}
\end{eqnarray}
\end{subequations}

\noindent
In particular, for the $B^{\alpha}$, the gauge-dependent part of $(\mathcal{O}_{SSD}^{-1})$ drops out, and we find
\begin{equation}
B_\alpha = - \frac{1}{m D^2} \left( D^{\beta}D_{\alpha}\frac{1}{2D^{2}} \right) j_\alpha =
- \frac{1}{m D^2} j_\alpha \,,
\end{equation}

\noindent
since $j^\alpha$ is transversal. The field-strength corresponding to this superpotential is found to be
\begin{equation}\label{CS}
W_{B}^{\,\,\alpha} = \frac{1}{2}D^{\beta}D^{\alpha}B_{\beta} = -\frac{1}{m}j^\alpha \,.
\end{equation}

\noindent
This is the supersymmetric version of the well known relation between the source and the field strength generated by this source in the Chern-Simons model~\cite{morosov}.

Substituting Eqs.~(\ref{src}) in the master Lagrangian, one obtains
\begin{eqnarray}\label{eq10a}
\mathcal{L}_{eff}&=&
-\frac{1}{4m^2} \left(k^\alpha - j^{\alpha}\right) 
\left(\frac{D^{\beta}D_{\alpha}}{D^2+m}\right)\left(k_\beta- j_\beta \right)
+\frac{1}{2 m^2} \left(k^\alpha - j^{\alpha}\right) 
\left(k_\alpha- j_\alpha \right) \nonumber\\
&&-\frac{1}{2m}j^{\alpha}\frac{1}{D^2}j_{\alpha} \, ,
\end{eqnarray}

\noindent
which is the same as~(\ref{eq10}).

\section{Some particular instances of the duality}\label{instances}

Having studied the correspondence between the SSD and the SMCS models in the presence of arbitrary matter currents, now we consider some interesting particular cases. 

The case (a) corresponds to the choice $j^{\alpha}=0$, the matter superfield interacting with the vector superfields only through the current $k^{\alpha}$. In this case, we can summarize our results as
\begin{subequations}\label{sdeq11}
\begin{align}\label{sdeq11a}
\mathcal{L}_{\rm SMCS}^{(a)}=&\frac{1}{2}W^{\alpha}W_{\alpha}
+\frac{m}{2}A^{\alpha}W_{\alpha}-\frac{\alpha}{4}(D^{\alpha}A_{\alpha})^2
+\frac{1}{2m}A^{\alpha}D^{\beta}D_{\alpha}k_{\beta}\nonumber\\
&+\frac{1}{2m^2}k^{\alpha}k_{\alpha}+\mathcal{L}_{M}(\Psi)~,\\
\mathcal{L}_{\rm SSD}^{(a)}=&-\frac{m}{2}f^{\alpha}\Omega_{\alpha}
-\frac{m^2}{2}f^{\alpha}f_{\alpha}+k^{\alpha}f_{\alpha}
+\mathcal{L}_{M}(\Psi)~,\label{sdeq11b}\\
\mathcal{L}_{eff}^{(a)}=&
-\frac{1}{4m^2}k^{\alpha}\Big(\frac{D^{\beta}D_{\alpha}}{D^2+m}\Big)k_{\beta}
+\frac{1}{2m^2}k^{\alpha}k_{\alpha}
+\mathcal{L}_{M}(\Psi)~.\label{sdeq11c}
\end{align}
\end{subequations}

\noindent
Comparing Eqs.~(\ref{sdeq11b}) and~(\ref{sdeq11c}), we see that the minimal interaction $k^\alpha f_\alpha$ induces in the $\Psi^\alpha$ effective Lagrangian a non-local interaction mediated by a massive degree of freedom, plus a contact interaction between the matter currents. Besides, we note that Eq.~(\ref{sdeq11a}) already contains the contact term, and the non-minimal interaction $A^{\alpha}D^{\beta}D_{\alpha}k_{\beta}$ between the matter and the Chern-Simons field is responsible for describing the non-local interaction in Eq.~(\ref{sdeq11c}.

Comparing the equations of motion for the matter, in Eqs.~(\ref{eqq:3}) and (\ref{eqq:13}), in this particular case we recognize the identification
\begin{eqnarray}\label{sdeq11cc}
f_{\gamma}=\frac{W_{\gamma}}{m}+\frac{k_{\gamma}}{m^2}\,,
\end{eqnarray}

\noindent
which have been found in~\cite{Ferrari:2006vy}, when studying the duality in the presence of a scalar matter superfield. Indeed, Eq.~(\ref{sdeq11b}) is analogous to the starting point of~\cite{Ferrari:2006vy}; here, however, the dual SMCS description, Eq.~(\ref{sdeq11a}), is simpler since it does not contain the field-dependent factor in front of the Maxwell term, due to the absence of quartic couplings.

Case (b) is a theory where matter interacts only through the current $j^{\alpha}$, i.e., $k^{\alpha}=0$. In this case we have
\begin{subequations}\label{sdeq12}
\begin{align}
\mathcal{L}_{\rm SMCS}^{(b)}=&\frac{1}{2}W^{\alpha}W_{\alpha}
+\frac{m}{2}A^{\alpha}W_{\alpha}-\frac{\alpha}{4}(D^{\alpha}A_{\alpha})^2
+j^{\alpha}A_{\alpha}+\mathcal{L}_{M}(\Psi)~,\label{sdeq12a} \\
\mathcal{L}_{\rm SSD}^{(b)}=&-\frac{m}{2}f^{\alpha}\Omega_{\alpha}
-\frac{m^2}{2}f^{\alpha}f_{\alpha}
-f^{\alpha}j_{\alpha}
-\frac{1}{2m}j^{\alpha}\frac{1}{D^2}j_{\alpha}
+\mathcal{L}_{M}(\Psi)~, \label{sdeq12b} \\
\mathcal{L}_{eff}^{(b)}=&
-\frac{1}{4m^2}j^{\alpha}\Big(\frac{D^{\beta}D_{\alpha}}{D^2+m}\Big)j_{\beta}
+\frac{1}{2m^2}j^{\alpha}j_{\alpha}
-\frac{1}{2m}j^{\alpha}\frac{1}{D^2}j_{\alpha}
+\mathcal{L}_{M}(\Psi)~.\label{sdeq12c} 
\end{align}
\end{subequations}

\noindent
Now, the minimal coupling of $j^\alpha$ to the Chern-Simons superfield corresponds to a non-local interaction mediated by a massive degree of freedom and a contact term for the $j^\alpha$, both similar to the ones in Eq.~(\ref{sdeq11c}), plus an additional Chern-Simons interaction (see discussion regarding Eq.~(\ref{CS})). Furthermore, if we substitute $j^\alpha=\frac{1}{2m}D^\beta D^\alpha g_\beta$ in Eqs.~(\ref{sdeq12a}) and~(\ref{sdeq12c}), we obtain in $\mathcal{L}_{eff}^{(b)}$ only the non-local interaction $\frac{1}{4m^2}g^{\alpha}\Big(\frac{D^{\beta}D_{\alpha}}{D^2+m}\Big)g_{\beta}$, which is consistent with the results discussed for the case (a). Finally, the explicit mapping between the (transversal parts of the) SSD and the SMCS superfields is given by
\begin{equation}
A_{\alpha}^{\perp}  =-f_{\alpha}^{\perp}-\frac{1}{mD^{2}}j_{\alpha}\,,
\end{equation}

\noindent
while their longitudinal parts are unrelated, as we pointed out earlier.

Case (c) corresponds to the choice $j^\alpha = k^\alpha$; in this case, from Eq.~(\ref{eq9a}), we decouple the matter from the self-dual superfield, so we end up with a free SSD superfield plus a Chern-Simons interaction between the matter currents. This is equivalent, from Eq.~(\ref{eq7a}), to a model including a local Thirring interaction, along with a special coupling $\left(\frac{D^2+m}{m} j^\alpha\right)A_\alpha$ to the Maxwell-Chern-Simons superfield. In other words, the coupling $\left(\frac{D^2+m}{m} j^\alpha\right)A_\alpha$ induces, in the effective action, the terms $-\frac{1}{2m^2}j^\alpha j_\alpha - j^\alpha \frac{1}{2mD^2}j_\alpha$.

Finally, the case (d) is the choice $j^{\alpha}=-\frac{1}{2m}D^{\beta}D^{\alpha}k_{\beta}$. In this case, we decouple matter and SMCS in Eq.~(\ref{eq7a}); from this, we immediately see that the effective Lagrangian for the matter contains only a Thirring $\frac{1}{2m^2}k^\alpha k_\alpha$ interaction. On the other side, from Eq.~(\ref{eq9a}), the same dynamics is described by a Self-Dual model with the coupling 
$\left(\frac{k^\alpha + \frac{D^2}{m} k^\alpha_\perp}{m} \right)f_\alpha$ between matter and self-dual superfields, plus a Thirring-like interaction $-\frac{1}{4m^3}k^\alpha D^\beta D_\alpha k_\beta$.

\section{The matter content}\label{matter}

As we discussed in the Introduction, the equivalence between the SD 
and the MCS theories is more simply established when these superfields
interact with a fermionic matter superfield $\Psi_{\alpha}$ throught
the currents $j^{\alpha}$ and $k^{\alpha}$, which, together with
the matter free Lagrangian $\mathcal{L}_{M}\left(\Psi\right)$, have
not been specified so far (except for the requirement that $j^{\alpha}$
is conserved, so that it can be coupled to a gauge superfield). In
this section, we write explicitly a free Lagrangian $\mathcal{L}_{M}\left(\Psi\right)$,
and investigate its physical degrees of freedom. We also point out
a simple choice for the current $j^{\alpha}$. Even if we cannot give a more physical motivation
for the introduction of such matter supermultiplet, at least we can
explicitly demonstrate that a sensible dynamics can be constructed
for such a model.

The component expansion of the fermionic superfield $\Psi^{\alpha}$
is given by
\begin{equation}
\Psi^{\alpha}=\psi^{\alpha}+\theta^{\alpha}b+i\theta_{\beta}b^{\beta\alpha}-\theta^{2}\varphi^{\alpha}\,,
\label{eq:1}
\end{equation}

\noindent
where $b^{\beta\alpha}$ is a symmetric bispinor (a three-dimensional vector field),
$b$ is a scalar, $\psi_{\alpha}$ and $\varphi_{\alpha}$ are three-dimensional spinors.
Since we want to couple this matter to a gauge superfield, these fields
are complex. The complex conjugate $\overline{\Psi}^{\alpha}$ can be written as 
\begin{equation}
\overline{\Psi}^{\alpha}=\overline{\psi}^{\alpha}+\theta^{\alpha}\overline{b}+i\theta_{\beta}\overline{b}^{\beta\alpha}-\theta^{2}\overline{\varphi}^{\alpha}\,.\label{eq:2}
\end{equation}

We choose to study the case when the matter interacts though the current $j^\alpha$, minimally coupled to the gauge superfield $A^\alpha$: the corresponding Lagrangian appears in Eq.~(\ref{sdeq12a}). Here we will focus only in the part of the action involving the $\Psi_{\alpha}$, i.e.,
\begin{equation}
S_{M} = \int d^5 z \, \mathcal{L}_{M}\left(\Psi\right) + \int d^5 z\, A^\alpha j_\alpha \,.
\end{equation}

\noindent
The proposed quadratic action for the matter superfield is given by
\begin{equation}
\int d^5 z \, \mathcal{L}_{M}\left(\Psi\right)\,=\,-\int d^{5}z\,\overline{\Psi}^{\alpha}\left(i\partial_{\alpha\beta}-M\, C_{\alpha\beta}\right)\Psi^{\beta}\,,\label{eq:5}
\end{equation}

\noindent
while the current $j^\alpha$ reads,
\begin{equation}
j^{\alpha}\,=\,\frac{i g}{2} D_{\beta} \left(\overline{\Psi}^{\alpha}\Psi^{\beta}+\overline{\Psi}^{\beta}\Psi^{\alpha}\right)\,.\label{eq:15}
\end{equation}

\noindent
This form of the matter current is obtained by the usual substitution
\begin{equation}
\partial_{\alpha\beta}\rightarrow\nabla_{\alpha\beta}=\partial_{\alpha\beta}-g\,D_{(\alpha}A_{\beta)}
\,,\label{eq:13}
\end{equation}

\noindent
in the quadratic Lagrangian $\mathcal{L}_{M}\left(\Psi\right)$, so that the action $S_M$ turns out to be invariant under the gauge transformations
\begin{equation}
\Psi^{\alpha}\rightarrow e^{igK}\Psi^{\alpha}\quad;\quad\overline{\Psi}^{\alpha}\rightarrow e^{-igK}\overline{\Psi}^{\alpha}\quad;\quad A_{\alpha}\rightarrow A_{\alpha}+D_{\alpha}K\,,\label{eq:14}
\end{equation}

\noindent
$K$ being a real scalar superfield. The coupling constant $g$ has mass dimension $1/2$, which in principle signals a super-renormalizable theory. 
By explicit computation, we verify that $D_{\alpha}j^{\alpha}=0$, as it should. Actually, for this particular form of the current $j^\alpha$, this conservation equation reduces to
\begin{equation}
i \partial_{\alpha\beta} \left( \overline{\Psi}^\alpha \Psi^\beta \right)\,=\,0\,.
\end{equation}

The component expansion of $S_{M}$ can be obtained with the help of the formulae in Appendix \ref{app}. The expansion of the $A^\alpha$ superfield in components reads
\begin{equation}
A^{\alpha}=\alpha^{\alpha}+\theta^{\alpha}a+i\theta_{\beta}a^{\beta\alpha}-\theta^{2}\beta^{\alpha}
\,,\label{eq:2a}
\end{equation}

\noindent
and, for simplicity, we will work in the Wess-Zumino gauge, so that $\alpha^\alpha = 0$ and $a=0$. The remaining vector field corresponds to the photon and the spinor to the photino. The matter action $S_M$ can be written, in terms of component fields, as follows,
\begin{equation}
S_M \, = \, S_M ^ {(1/2)} + S_M ^ {(1)} + S_M ^ {\textrm{(int)}}\,
\end{equation}

\noindent
where
\begin{equation}
S_M ^ {(1/2)}\,=\,\int d^{3}x \,
\left[ \overline{\varphi} \left(i\,\gamma^a\partial_a-M\right) \psi
+\overline{\psi} \left(i\,\gamma^a\partial_a-M\right) \varphi \right]\,,\label{eq:7}
\end{equation}
\begin{equation}
S_M ^ {(1)}\,=\,-\int d^{3}x  \,\left[
\frac{1}{2} \varepsilon^{abc}\overline{b}_{a}\partial_{b}b_{c}
+\frac{M}{2}\,\overline{b}^{a}b_{a}
+\overline{b} \partial^a b_a + {b} \partial^a \overline{b}_a - 2 M \overline{b} b
\right]\,,\label{eq:8}
\end{equation}

\noindent
and
\begin{align}
S_M ^ {\textrm{(int)}} \, = \, g \int d^5 z &\left[ 
a_a \left( \overline{\psi} \gamma^a \varphi + \overline{\varphi} \gamma^a \psi \right) 
-\frac{1}{2} a_a \varepsilon^{abc}\partial_b \left( \overline{\psi} \gamma_c \psi \right) \right. \\
&\left. 
+\overline{b} (\psi \beta) + b (\overline{\psi} \beta)
+ \frac{i}{2} b_a \left( \overline{\psi} \gamma^a \beta \right)
+ \frac{i}{2} \overline{b}_a \left( {\beta} \gamma^a \psi \right) \right] \,. \nonumber
\end{align}

\noindent
In writing these equations, we have used that $\left(i\,\gamma^a\partial_a-M\right)_{\alpha\cdot}^{\cdot\beta}=\left(i\,\partial_{\alpha\cdot}^{\cdot\beta}-\delta_{\alpha\cdot}^{\cdot\beta}M\right)$, and the relation between a bispinor and a vector $X^{\alpha\beta} = 1/2 (\gamma^a)^{\alpha\beta} X_a$ (see Appendix); latin indices run from $0$ to $2$. The $b$ is an auxiliary superfield, that can be eliminated by means of its equation of motion, and we obtain
\begin{equation}
S_M ^ {(1)}\,=\,-\int d^{3}x  \,\left[
\frac{1}{2} \varepsilon^{abc}\overline{b}_{a}\partial_{b}b_{c}
+\frac{M}{2}\,\overline{b}^{a}b_{a}
- \frac{1}{2M} \left( \partial^a \overline{b}_a  \right) \left( \partial^b b_b  \right)
\right]\,,\label{eq:8b}
\end{equation}

The action $S_M ^ {(1)}$ corresponds to a kind of gauge-fixed Chern-Simons
theory, with a Proca mass term. There is no
gauge symmetry associated to the vector field $b_{\mu}$, thus the
\emph{complex} $b_{a}$ has four propagating degrees of freedom
with mass $M$, as can be seen by its propagator in momentum space,
\begin{equation}
\Delta_{ab}\left(k\right)\,=\,\frac{i}{k^{2}+M^{2}-i\varepsilon}
\left[i\varepsilon_{abc}k^{c}+M g_{ab}\right]\,.\label{eq:12}
\end{equation}

\noindent
This propagator is clearly not transversal, as it should be.

From the action $S_M ^ {(1/2)}$,
one obtains the usual equations of motion for spinors in three dimensions,
\begin{equation}
\left( i\gamma^a\partial_a - M \right) \psi=0\quad;\quad 
\left( i\gamma^a\partial_a - M \right)\varphi=0\,.\label{eq:10}
\end{equation}

\noindent
Since each real spinor has one on-shell degree of freedom, the
action  describes the propagation of \emph{four} fermionic
degrees of freedom. Notice however the mixing between the spinors $\psi_{\alpha}$ and $\varphi_{\alpha}$ already in the quadratic part of the action. 

The propagator in Eq.~(\ref{eq:12}) has components with indefinite metric, as can be seen by simple inspection. Not surprisingly, the same problem appears in the fermionic sector: if we try to disentangle the $\psi^\alpha$ and $\varphi^\alpha$ fields by diagonalizing the quadratic action in Eq.~(\ref{eq:7}), the new fermionic kinetic terms end up with opposite signs, also indicating and indefinite metric in the space of quantum states. This is not an unusual feature in quantum field theory. In  fact,
the presence of indefinite metric actually permeates the quantization of any gauge theory; in those cases the quantization has to be suplemented by selection  rules to extract
physically relevant results. We intend to come back to this issue in a future publication.

\section{Summary}

Let us summarize our results. We have studied the dual equivalence of the supersymmetric self-dual and Maxwell-Chern-Simons theories, coupled to a fermionic matter superfield. We have shown these models to be equivalent at the classical level, by looking at their equations of motion, which actually provides us with a mapping between fields and currents of both theories. At the quantum level, their equivalence follows from the equality of the effective generating functional they induce for the matter. The duality holds in presence of matter currents $j^\alpha$ and $k^\alpha$ that are quite arbitrary, the only requirement being that $D_\alpha j^\alpha=0$ so that $j^\alpha$ can be coupled to the gauge superfield. 

The duality in the presence of such a non-minimal matter superfield is much simpler than with an usual scalar matter superfield, discussed in~\cite{Ferrari:2006vy}. We have also shown how a sensible dynamics can be given to such an unusual supermultiplet, as well as how they can be coupled to the gauge superfield.

As a final remark, we comment on the possible extension of our results to the noncommutative case. 
We remark that, for a scalar matter superfield, no such extension was possible using the gauge embedding method~\cite{Ferrari:2006vy}.
In the present case, all manipulations needed to verify the duality were done without specifying the real nature of the currents $j_{\alpha}$ and $k_{\alpha}$, that can be treated as composite fields. We recall the property of the Moyal-Groenewald *-product that, inside an integral, one *-product in a monomial of fields can be replaced by an usual product, i.e.~\cite{reviews},
\begin{equation}
\int d^3 x \, \phi_1 * \phi_2 * \phi_3 * \cdots * \phi_n = \int d^3 x \, \phi_1 \left( \phi_2 * \phi_3 * \cdots * \phi_n\right)\,.
\end{equation}

\noindent
That means we can generalize the master action, substituting all usual products by Moyal-Groenewald products, and we end up with
\begin{align}\label{nc1}
S_{\rm master}^{\rm (*)}=\int d^5z & \left[-\frac{m^2}{2}f^{\alpha}f_{\alpha}+m~f^{\alpha}W_{\alpha}
+\frac{m}{2}A^{\alpha}W_{\alpha} \right. \nonumber\\
&\left. + k_{*}^{\alpha}f_{\alpha}+j_{*}^{\alpha}A_{\alpha}+\mathcal{L}_{M}(\Psi) \right]~,
\end{align}

\noindent
where the *-product appears only inside the currents $j$ and $k$. In this way, the proof of equivalence between the SSD and SMCS theories follows as in the previous sections. Notice, however, that these are not full-fledged noncommutative SSD or SMCS theories, since the *-product only affects the matter currents. The difficulties in studying the equivalence between noncommutative SSD and SMCS theories cannot be solved by the methods presented in this paper; see however \cite{Gomes:2008pi}.

\vspace{1cm}
{\bf Acknowledgments.} This work was partially supported by Conselho Nacional de Desenvolvimento Cient\'{\i}fico e Tecnol\'{o}gico (CNPq) and Funda\c{c}\~{a}o de Amparo \`{a} Pesquisa do Estado de S\~{a}o Paulo (FAPESP). A.C.L. is supported by FAPESP project No. 2007/08604-1.

\appendix
\section{Conventions on the three-dimensional $\gamma$ matrices}\label{app}

In this paper, we use the conventions of~\cite{Gates:1983nr}, where
the $C^{\alpha\beta}$ tensor, used to contract spinorial indices,
is purely imaginary. This fact requires care when raising and lowering
indices, since 
$\overline{\psi}^{\alpha}=\left(C^{\alpha\beta}\psi_{\beta}\right)^{*}
=-C^{\alpha\beta}\overline{\psi}_{\beta}$.
Thus, for example, while $\theta^{\alpha}$ is assumed to be real,
$\theta_{\alpha}$ is imaginary, and $\theta^{2}=\frac{1}{2}\theta^\alpha \theta_\alpha$ is real.
On the other hand, for a complex spinor $\psi_{\alpha}$, we have,
for example, $\left(\theta^{\alpha}\psi_{\alpha}\right)^{*}=-\theta^{\alpha}\overline{\psi}_{\alpha}$. 

Three-dimensional $\gamma_{\,\,\beta}^{\alpha}$ matrices are required to satisfy
\begin{equation}
\left\{ \gamma^{a},\gamma^{b}\right\} _{\,\,\beta}^{\alpha} = 2\delta_{\,\,\beta}^{\alpha}\eta^{ab}\,,\label{eq:ap1}
\end{equation}

\noindent
where $\eta^{ab}=\mbox{diag}\left(-++\right)$; latin indices run from 0 to 2. 
We require the
$\gamma_{\,\,\beta}^{\alpha}$ matrices to be \emph{traceless}, and
remember that the tensor $C_{\alpha\beta}$ used to raise and lower
spinor indices can be written in matrix representation as $\left[C_{\alpha\beta}\right]=-\left[C^{\alpha\beta}\right]=\sigma^{2}$.
Here, $\sigma^{i}$ denote the standard Pauli matrices. One choice
for $\gamma^{a}$ satisfying such requirements is
\begin{equation}
\left(\gamma^{a}\right)_{\,\,\beta}^{\alpha}=
\left(-\sigma^{2},-i\sigma^{1},i\sigma^{3}\right)\,.\label{eq:ap2}
\end{equation}

\noindent
By lowering and raising spinor indices, we have also
\begin{align}
\left(\gamma^{a}\right)_{\alpha\beta} & =\left(\unity,\sigma^{3},\sigma^{1}\right)\,,\label{eq:ap3}\\
\left(\gamma^{a}\right)^{\alpha\beta} & =\left(-\unity,\sigma^{3},\sigma^{1}\right)\,.\label{eq:ap4}
\end{align}

The $\gamma$ matrices are used to pass from the bispinor representation
for a vector field and the more usual ``vectorial'' representation,
by means of the relations
\begin{equation}
X^{a}=\left(\gamma^{a}\right)^{\alpha\beta}X_{\alpha\beta}\quad;\quad X^{\alpha\beta}=\frac{1}{2}\left(\gamma^{a}\right)^{\alpha\beta}X_{a}\,.\label{eq:ap8}\end{equation}
From Eq.\,(\ref{eq:ap8}), one obtains
\begin{equation}
\partial_{a} = \frac{1}{2}\left(\gamma^{a}\right)^{\alpha\beta} \partial_{\alpha\beta}\quad;\quad\partial_{\alpha\beta} =\left(\gamma^{a}\right)_{\alpha\beta}\partial_{a}\,.\label{eq:ap9}
\end{equation}

\noindent
The particular normalization chosen in Eqs.\,(\ref{eq:ap8}-\ref{eq:ap9})
is such that the relation
\begin{equation}
\partial^{\alpha\beta}\partial_{\beta\sigma}=\delta_{\,\,\sigma}^{\alpha}\Box\label{eq:ap10}
\end{equation}

\noindent
holds, as in~\cite{Gates:1983nr}. On the other hand, one has
\begin{equation}
X^{\alpha\beta}X_{\beta\sigma}=\frac{1}{4} \delta_{\,\,\sigma}^{\alpha}X^a X_a \,,\label{eq:ap11}
\end{equation}

\noindent
for the vector $X^a$.


\begin{thebibliography}{99}
\bibitem{foot1} 
One can easily verify that the inclusion of the gauge fixing term $-\frac{\alpha}{4}(D^{\alpha}A_{\alpha})^2$ in $\mathcal{L}_{\rm SMCS}$ does not modify the equation of motion for the transversal part of $A^\alpha$. Indeed, using the transversal projector in~(\ref{eq:project}), one has
\begin{equation*}
D^{\beta}D^{\alpha}\frac{1}{2D^{2}}\left[-\frac{\alpha}{2}D_{\beta}\left(D^{\gamma}A_{\gamma}^{\parallel}\right)\right]=-\frac{\alpha}{4D^{2}}D^{\beta}D^{\alpha}D_{\beta}\left(D^{\gamma}A_{\gamma}^{\parallel}\right)=0
\end{equation*}
due to the well known identity $D^{\beta}D^{\alpha}D_{\beta}=0$.
\end{thebibliography}
\end{document}